\documentclass[twocolumn,prl,amssymb,nofootinbib,nokeywords,amsmath,showpacs]{revtex4}
\usepackage{epsfig}
\usepackage{latexsym}
\usepackage{graphicx}
\usepackage{amssymb}
\usepackage{subfigure}
\usepackage{hyperref}
\usepackage{verbatim}
\newcommand{\sigmav}{\langle \sigma_a v \rangle}
\newcommand{\sigmaiv}{\langle \sigma_{i j \rightarrow \chi\chi} v \rangle}
\newcommand{\sigmaivs}{\langle \sigma_{i j \rightarrow ? \chi} v \rangle}
\newcommand{\SM}{ {\scriptscriptstyle \rm SM}}
\newcommand{\ManS}{\mathsf s}

\begin{document}

\DeclareGraphicsExtensions{.pdf,.png,.gif,.jpg,.eps}

\title{Hidden Hot Dark Matter as Cold Dark Matter}

\author{Kris Sigurdson}
\email{krs@physics.ubc.ca}
\affiliation{Department of Physics and Astronomy, University of British Columbia, Vancouver, BC V6T 1Z1, Canada}

%------------------------------------------------------------------------------

\begin{abstract}
We show that hidden hot dark matter, hidden-sector dark matter with interactions that decouple when it is relativistic, is a viable dark matter candidate provided it has never been in thermal equilibrium with the particles of the standard model.  This hidden hot dark matter may reheat to a lower temperature and number density than the visible Universe and thus account, simply with its thermal abundance, for all the dark matter in the Universe while evading the typical constraints on hot dark matter arising from structure formation.  We find masses ranging from $\sim\!\!3$~keV to $\sim\!\!10$~TeV.  While never in equilibrium with the standard model, this class of models may have unique observational signatures in the matter power spectrum or via extra-weak interactions with standard model particles. 
\end{abstract}

\keywords{Dark Matter, Cosmology}

%------------------------------------------------------------------------------
% User-supplied List of keywords.

\pacs{95.35.+d, 98.80.Cq, 14.80.-j}

\maketitle
%\section{Introduction}\label{sec:intro}
\noindent \emph{Introduction} --
Dark matter is a known unknown.  While we know its abundance \cite{WMAP} and some of its cosmological and astrophysical properties (see, e.g., Ref.~\cite{D'Amico:2009df})
its fundamental nature remains elusive.  Viable candidates, that may also play a role in the solution of  other pressing issues in physics, have been proposed including weakly interacting massive particles (WIMPs) \cite{WIMPs}, and axions \cite{axions}.
Although they have drastically different properties both would have been relatively\emph{ cold} dark matter (CDM) particles (relative to their mass) when first produced.  They would have moved only a short distance, cosmologically speaking, before their initial momentum redshifted away.

On the contrary, while standard model (SM) neutrinos have mass \cite{neutrinomass}, they fail as a candidate for dark matter because they decoupled from the primordial plasma at a high temperature relative to their mass.  They are \emph{hot} dark matter (HDM) particles that stream through the Universe 
at velocities $v\simeq c$ and travel huge cosmological distances before slowing down.   This free-streaming would erase small-scale density fluctuations in violent conflict with observations \cite{wdm}.   Moreover, to account for a cosmological dark matter density of $\Omega_d h^2 \simeq 0.11$ \cite{WMAP} a single Dirac (Majorana) fermion that decoupled while relativistic must have a mass $\simeq\!\!50$~eV ($\simeq\!\!100$~eV) if it decoupled prior to the QCD and electroweak phase transitions, and a \emph{lower} mass if (like neutrinos) it decoupled later.  As the Tremaine-Gunn bound on the phase space density of dark matter requires $m \gtrsim 300$~eV \cite{TG}, it appears that an interacting species once in thermal equilibrium that decouples when relativistic, like the SM neutrinos, can only account for a minor fraction of the dark matter in the Universe.

However, there is an important caveat to this conclusion.  It assumes that dark matter was, at some point in the history of the Universe, in thermal equilibrium with \emph{standard model particles}. This need not be the case if dark matter exists in a hidden sector (HS). In fact, dark-matter could have been thermalized with \emph{hidden sector particles} at a HS temperature less than the SM temperature.\footnote{A different possibility, one example being the Dodelson-Windrow sterile neutrino \cite{Dodelson:1993je}, is non-thermal dark matter that \emph{never} equilibrates with \emph{any} species. This alternative is not our focus here.} This may occur straightforwardly, for instance, if reheating after inflation proceeds more efficiently to the SM sector than the HS.\footnote{The relative temperature could also  be reduced if $g^{s}$ decreases by a huge factor before electroweak but after hidden decoupling.}  Hidden sectors have long been imagined  \cite{Mirror},  and HS dark matter candidates that freeze out when non-relativistic have recently received attention \cite{HCDM}.  Potential implications of extra HS interactions have also been considered (see, e.g., Refs.~\cite{HiddenInteractions}).  

In this \emph{Letter} we show that hidden hot dark matter (HHDM), a thermalized HS species with interactions that decouple while relativistic, is a candidate for the all of the dark matter in our Universe.   This alternative class of dark matter candidates has a thermal relic abundance, significantly different particle properties than standard thermal relics, and might have extra-weak interactions with the SM.

%\section{Mass and Relic Density of HHDM}\label{sec:density}
\vspace{0.04in}
\noindent {\em Mass and Relic Density} -- We assume that dark matter is a particle $\chi$ of mass $m_{\chi}$ which freezes out when it is relativistic.  This analysis is for a fermion, but appropriately adapted results  hold for bosons.  A fermion with $d_{\chi}$ internal degrees of freedom in thermal equilibrium at a temperature $T_{\chi}$ has a number density
\begin{align}
n_{\chi} = d_{\chi}\frac{3 \zeta(3)}{4\pi^2} T_{\chi}^3 = d_{\chi}\frac{3 \zeta(3)}{4\pi^2} \frac{g^s(T)}{g^s_f}\xi^3 T^3\, ,
\label{eq:nchi}
\end{align} 
where $T$ is the SM temperature, $g^s(T)$ is the effective number of entropy degrees of freedom coupled to the SM, and $\xi \equiv T_{\chi}/T$ is the dark-to-SM temperature ratio at the temperature $T_{\chi}$.   When HS interactions of $\chi$ particles decoupled at temperature $T_{\chi f}$ we have $\xi_f\equiv T_{\chi f}/T_f$.  Here $\zeta(3) \simeq 1.202$ and for Dirac fermion \mbox{$d_{\chi} = 4$} counting both particles and antiparticles (for a Majorona fermion $d_{\chi} = 2$). 
If $\chi$ particles are dark matter they are now non-relativistic and have an energy density 
\begin{align}
\rho_{\chi0} \equiv \Omega_d \rho_{\rm crit} = m_{\chi}  n_{\chi f} \left(\frac{a_0}{a_f}\right)^{-3}  \, ,
\end{align}
where $n_{\chi f}$ is the $\chi$ number density at freeze-out, $a_{f}$ the scale factor at freeze-out and $a_0$ the scale factor at present.  This means that $\chi$ particles must have a mass
\begin{align}
m_{\chi} =  40\left(\frac{\Omega_d h^2}{0.11}\right)\!\!\left(\frac{4}{d_{\chi}}\right)\!\!\left(\frac{g^s_f}{g^s_0}\frac{3.91}{83}\right)\!\!\left(\frac{0.1}{\xi_{f}}\right)^3\,\,{\rm keV}
\end{align}
in order to be dark matter.   Here $g^s_f$ and $g^s_0$ are the effective number of entropy degrees of freedom in the Universe at $\chi$ decoupling and at present respectively.

We emphasize that as $\chi$ particles decouple when relativistic, the dark-matter density is determined by the thermal abundance of a relativistic species, and is expressed in terms of the temperature and thermal history of the Universe.  
Unlike non-relativistic freeze-out, it is independent of the interaction cross section except for requiring it be small enough to allow relativistic decoupling.
Instead we find, at fixed $\Omega_d h^2$, there is a unique relationship between mass $m_{\chi}$ and the dark-to-SM temperature-ratio $\xi_{f}$. 

As it is cold relative to the SM plasma, we show below the lower limit on the dark-matter mass from structure-formation are slightly relaxed and $m_{\chi} \simeq 3$~keV is viable.  
For small $\xi_{f}$, this HDM might still be very heavy and, as discussed below, may have masses as large  as $m_{\chi} \sim 10$~TeV for the class of models we discuss.
On the low mass end HHDM acts like warm dark matter and alters the dark-matter power spectrum on relatively large length scales where it is sensitive to constraints from cosmology or observations of substructure.  Furthermore, while the thermal history of dark matter is determined by HS interactions this class of dark-matter candidates may also couple to the SM with extra-weak interactions.  Extra-weak interactions can result, for instance, in the decay of dark matter to SM particles on timescales longer than the age of the Universe and imprint a characteristic energy scale $ 3 \,{\rm keV} \lesssim m_{\chi} \lesssim 10$~TeV on any SM decay products.

\begin{figure*}[t]
\includegraphics[width=18 cm]{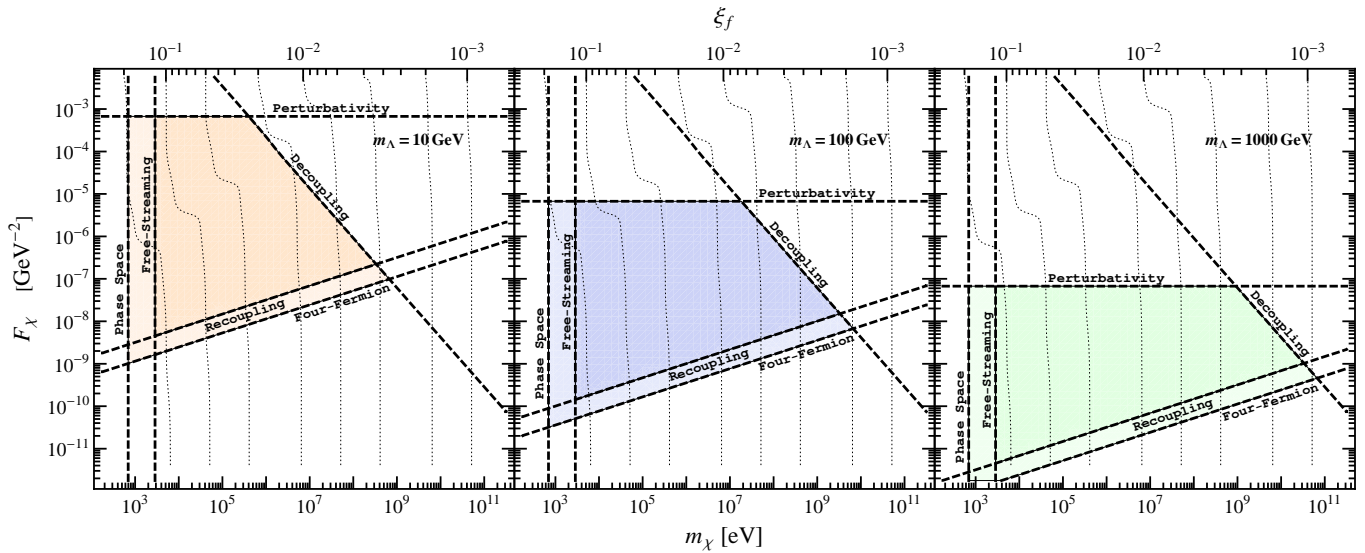}
  \caption{The viable HHDM ($m_{\chi}$,$F_{\chi}$) parameter space for $m_{\Lambda} = 10$~GeV (left), $m_{\Lambda} = 100$~GeV (middle), and  $m_{\Lambda} = 1000$~GeV (right) is shown for  $\Omega_d h^2 = 0.11$, $d_{\chi}=4$, $N_{\rm ch}=1$, $g^s_{Dr}=g^s_{Df}$, and a standard model thermal history in the SM sector.  The darker shading  shows the allowed region if the free-streaming constraint and recouping constraint are imposed (the latter with ${\cal E}_r = 1000$).  The lighter shading shows how the allowed region expands if the phase space density and four-fermion constraints are used instead.  The top scale, and the associated dotted contours, shows the dark-to-SM temperature ratio at freeze-out $\xi_f$.  For fixed $\xi_f$ models with lower $F_{\chi}$, that freeze out prior to the QCD phase transition, require a larger $m_{\chi}$ to achieve a dark matter density $\Omega_{d} h^2 = 0.11$ compared to models that freeze out later.   Viable corners of parameter space exist over the range $ 10^{-4}\, {\rm GeV}  \lesssim m_{\Lambda} \lesssim 10^{11}$~GeV , while the maximum $\chi$ mass allowed is $m_{\chi} \sim 10$~TeV near $m_{\Lambda} \simeq 150$~TeV.}
  
\label{fig:constraints}
\end{figure*}

%\section{HHDM Decoupling }\label{sec:model}
\vspace{0.04in}
\noindent \emph{Decoupling} --
We discuss here our physical assumptions and determine constraints on the decoupling of $\chi$ from the HS plasma so that it is viable HHDM.  Instead of focusing on a specific theory, we imagine that the annihilation of $\chi$ particles is mediated by a set of particles with typical coupling $\lambda$ and high-energy mass scale $m_{\Lambda}$.   For instance, if $\chi$ were similar to a SM neutrino then $\lambda$ would be a typical weak-scale coupling, and $m_{\Lambda}$ would be close to the mass of the $W$ boson.  While not completely general  this setup encompasses a wide variety of models and has a natural analogy with the weak interactions of SM neutrinos.

If $\chi$ particles annihilate via the exchange of a relatively massive particle we can estimate the annihilation cross section on kinematic and dimensional grounds. If the $\chi$ particles have momentum $p \gg m_{\Lambda}$ then the annihilation cross section goes like $\sigma_a \sim \lambda^4/\ManS$ where $\ManS$ is the Lorentz-invariant Mandelstam variable.  However, for particles with momentum $ m_{\chi} \ll p \ll m_{\Lambda}$, like for SM weak interactions, the propagator of the mediator is dominated by $m_{\Lambda}$ and $\sigma_a \sim G_{\Lambda}^2 \ManS$, where $G_{\Lambda} \equiv  \lambda^2/m_{\Lambda}^2$ is an effective four-fermion coupling constant.   For $p \lesssim m_{\chi}$ we expect $\sigma_a \sim G_{\Lambda}^2m_{\chi}^2$, but further kinematic suppression is possible depending on the structure of the effective interaction.  

For hot production we are in the intermediate regime where $\sigma_{a} \sim G_{\Lambda}^2 \ManS$ and so, on dimensional grounds, the thermalized cross section (times velocity) goes like $\sigmav \sim G_{\Lambda}^2 T_{\chi}^2$ where $T_{\chi}$ is the $\chi$ particle temperature.   In viable models
$T_{\chi}$ must be less than the SM temperature $T$.
In terms of the temperature ratio $\xi$ the total thermalized annihilation rate is
\begin{align}
\Gamma_{a} = n_{\chi}\sigmav  \equiv F_{\chi}^2 T_{\chi}^5 = F_{\chi}^2 \xi^5 T^5 \, .
\label{eq:gamma}
\end{align}
where we have used $n_{\chi} \propto T_{\chi}^3$ for a relativistic species.   We can write the total thermalized cross section as
%\begin{align}
$\sigmav \equiv N_{\rm ch} G_{\Lambda}^2 T_{\chi}^2$,
%\end{align}
where $N_{\rm ch}$ is the effective number of channels involved in number-changing interactions.  This factor accounts for the number of channels via which $\chi$ can annihilate, differences in the true rates of the relevant processes, and factors arising from the thermal rate calculation.
The thermalized coupling $F_{\chi}$ and effective coupling $G_{\Lambda}$ are related by \mbox{$F_{\chi} \simeq 0.6 \sqrt{N_{\rm ch}}\sqrt{d_{\chi}/4} \, G_{\Lambda}$}.

The freeze-out temperature $T_f$ is the \emph{standard model} sector temperature below which
\begin{align}
\Gamma_a |_{{T_{\chi f}}=\xi_f T_f} \leq H |_{T_{f}}\, 
\end{align}
and number-changing interactions become ineffective compared to expansion.
In a model with HS degrees of freedom the Hubble rate during radiation domination is
\begin{align}
H &= \frac{2\pi^{3/2}}{3\sqrt{5}}\sqrt{g(T)  + \xi_h^4 g_{h}(\xi_{h} T)}\frac{T^2}{M_{\rm Pl}} \, ,
\end{align}
where $g(T)$ and $g_h(\xi_h T)$ are the effective number of degrees of freedom in the SM sector and HS respectively, and $\xi_h \equiv T_h/T$ is the hidden-to-SM temperature ratio (equal to $\xi$ for $T \gtrsim T_f$).  In viable models we have $\xi_f \lesssim 0.1$ and thus $g_h \xi_h ^4 \lll  g$ (unless $g_h$ is very large).  This means the SM sector dominates the expansion of the Universe and we recover the standard form
\begin{align}
H \simeq \frac{2\pi^{3/2}}{3\sqrt{5}}\sqrt{g(T)}\frac{T^2}{M_{\rm Pl}}\simeq 1.66\sqrt{g(T)}\frac{T^2}{M_{\rm Pl}} \, .
\label{eq:hubble}
\end{align}
We use this approximation in what follows.

Using Eqs.~(\ref{eq:gamma}) and (\ref{eq:hubble}) we find a freeze-out temperature 
\begin{align}
T_f \simeq 16.5 \, \left(\frac{g_f}{11}\right)^{1/6}\!\!\left(\frac{\xi_f}{0.1}\right)^{-5/3}\!\!\left(\frac{F_{\chi}}{10^{-4} {\rm GeV}^{-2}}\right)^{-2/3}\, {\rm MeV} \, 
\label{eq:Tf}
\end{align}
We must ensure freeze-out occurs when $\chi$ particles are relativistic.
Requiring that $T_{\chi f} = \xi_{f} T_{f} \gtrsim 3m_{\chi}$ leads to the constraint
\begin{align}
\left(\frac{F_{\chi}} {10^{-4} \,{\rm GeV}^{-2}}\right)\!\!\left(\frac{m_{\chi}}{2\,{\rm MeV}}\right)^{7/6} 
 \lesssim \left(\frac{{\cal C}_m}{0.4}\right)^{-1/3}\!\!\left(\frac{g_{f}}{83} \right)^{1/4} 
 \end{align}
on $F_{\chi} m_{\chi}^{7/6}$, where ${\cal C}_m \equiv (\Omega_{d} h^2/d_{\chi})(g_f^s/g_0^s)$.  This is shown in Fig.~(\ref{fig:constraints}) as the curve labeled \emph{Decoupling}.  

We have assumed here an effective four-fermion model is a faithful approximation to interactions for $T_{\chi} \lesssim m_{\Lambda}$.   This imposes consistency constraints on the model for a given high-energy scale $m_{\Lambda}$.  
Requiring $T_{\chi f} \lesssim  m_{\Lambda}/5$ we find the constraint
\begin{align}
\left(\frac{F_{\chi}m_{\Lambda}^{3/2}}{10^{-7}\,{\rm GeV}^{-1/2}}\right)\!\!\left(\frac{m_{\chi}}{{\rm MeV}}\right)^{-1/3}  \gtrsim \frac{7}{2} \left(\frac{{\cal C}_m}{0.75}\right)^{-1/3}\!\!\left(\frac{g_f}{99}\right)^{1/4}
\end{align}
on $F_{\chi}m_{\Lambda}^{3/2}m_{\chi}^{-1/3}$.  This is shown in Fig.~{\ref{fig:constraints} as the curve labeled \mbox{\emph{Four-Fermion}}.   If $\lambda < 1/3$, so the dimensionless coupling $\lambda$ of the high-energy theory is perturbative at relevant energy scales, we find an upper limit
\begin{align}
F_{\chi} m_{\Lambda}^2 \lesssim \sqrt{\frac{\zeta(3)}{3}} \frac{\sqrt{N_{\rm ch} d_{\chi}}}{6\pi} \simeq 0.067 \sqrt{N_{\rm ch}}\sqrt{\frac{d_{\chi}}{4}}\, .
\end{align}
This is shown in Fig.~{\ref{fig:constraints} as the curve labeled \mbox{\emph{Perturbatvity}}. These consistency constraints might be relaxed or altered if a model is specified in detail.

%\section{HHDM Recoupling}
\vspace{0.04in}
\noindent \emph{Recoupling} --
If the $\chi$ number density is  given by its thermal distribution function the HS interactions discussed above (at least minimaly without extra high-energy interactions) must be strong enough that $\chi$ particles can thermalize in the early Universe after HS reheating.  This places a potential \emph{recoupling} constraint on the model.

At very high momentum $p \gg m_{\Lambda}$ we assume the number-changing cross section scales as $\sigma_a \sim {\lambda}^4/\ManS$. This means for $T \gtrsim m_{\Lambda}$ the annihilation rate is approximately
\begin{align}
\Gamma_a \simeq F_{\chi}^2 m_{\Lambda}^4 T_{\chi} \, .
\end{align}
We thus expect the ratio $\Gamma_a/H$ peaks near $T_{\chi} \sim m_{\Lambda}$.
For these interactions to thermalize $\chi$ particles we require $T_{\chi r} = \xi_{\chi} T_r$, the dark-matter temperature when
\begin{align}
\Gamma_a |_{T_r} \geq H |_{T_r} \,
\end{align}
is first satisfied, to be greater than $m_{\Lambda}$.  In particular we require $T_{\chi r} \gtrsim {\cal E}_r m_{\Lambda}$,
where ${\cal E}_r$ is a buffer parameter to allow for the time it may take the HS to thermalize from its initial distribution.  In what follows we set ${\cal E}_r \simeq 1000$, and the limit on $F_{\chi} m_{\Lambda}^{3/2}m_{\chi}^{-1/3}$ is
\begin{align}
\left(\frac{F_{\chi}m_{\Lambda}^{3/2}m_{\chi}^{-1/3}}{10^{-5}\,{\rm GeV}^{-5/6}}\right)\gtrsim \left(\frac{{\cal C}_m { \cal R}^s_{r,f}}{0.77}\right)^{-1/3}\!\!\!\!\sqrt{\frac{{\cal E}_r}{10^3}}\left(\frac{g_r}{107}\right)^{1/4} ,
\end{align}
where ${\cal R}^s_{r,f} \equiv  (g^s_r/g^s_f)(g^s_{Df}/g^s_{Dr})$. This is shown in Fig.~{\ref{fig:constraints} as the curve labeled \mbox{\emph{Recoupling}}, and applies if no other thermalization mechanisms are present in the HS.

%\section{HHDM Structure Formation}
\vspace{0.04in}
\noindent \emph{Structure Formation} --
For low masses $\chi$ particles appear to us like warm rather than cold dark matter and appropriately adapted constraints from cosmological structure formation apply.   After last interacting $\chi$ particles stream unimpeded through the Universe with $v\simeq c$ until their momentum (which redshifts as $p \propto a^{-1}$) is small compared $m_{\chi}$.  This occurs at a HS temperature $T_{\chi{\rm NR}} \simeq m_{\chi}/3.15$ and a SM-sector temperature $T_{\rm NR} \simeq (m_{\chi}/(3.15\,\xi_f)) (g^s_{f}/g^s_{\rm NR})^{1/3}$ .  We estimate the comoving free-streaming length is (see, e.g., Ref.~\cite{Kolb:1990vq})

\begin{align}
\lambda_{FS} \simeq \left(\frac{82}{9}+\ln\left[\left(\frac{42.5\, {\rm kpc}\cdot 0.137}{\lambda_{FS_{R}}\cdot \Omega_m h^2}\right)\left(\frac{\tilde{g}_{\rm NR}}{\tilde{g}_0}\right)^{1/6}\right]\right)\lambda_{{FS}_{\rm R}}
\end{align}
where 
\begin{align}
\lambda_{{FS}_{\rm R}} \simeq 42.5 \,{\rm kpc} \left(\frac{{\rm keV}}{m_{\chi}}\right)^{4/3}\!\!\left(\frac{4\cdot \Omega_d h^2}{d_{\chi} \cdot 0.11}\right)^{1/3}\!\!\left(\frac{\tilde{g}_{\rm NR}}{\tilde{g}_0}\right)^{1/6}.
\label{eq:lfs}
\end{align}
and we define, here and elsewhere, $\tilde{g} \equiv g^s \cdot (g^s/g)^3$.  In Eq.~({\ref{eq:lfs}), $\lambda_{{FS}_{\rm R}}$ is the distance the particle free-streams when ultrarelativistic, and $\lambda_{FS}$ includes the distance traversed when semi-relativistic prior to matter-radiation equality.   
As $\chi$ particles stream isotropically features in the initial density field smaller than $\lambda_{FS}$ are erased.   Limits from measurements of the matter power spectrum indicate $\lambda_{FS} \lesssim 110 \, {\rm kpc}$ which corresponds, up to logarithmic corrections, to the limit $m_{\chi} \gtrsim 2.9$~keV.  This is shown in Figure.~{\ref{fig:constraints} as the curve labeled \mbox{\emph{Free-Streaming}}. 

Another limit on $m_{\chi}$ arises from the Tremaine-Gunn bound \cite{TG}.   The maximum of the coarse grained $\chi$ phase space density (PSD) in a galaxy today with core radius $r_{c}$ and velocity dispersion $\sigma$ must be less then the maximum of the microscopic PSD (the Fermi-Dirac distribution) after interactions freeze out.  This leads to the constraint
\begin{align}
m_{\chi} \gtrsim \left(\frac{9\sqrt{2\pi}\hbar^3}{d_{\chi} G_N \sigma r_{c}^2}\right)^{1/4}\!\!\!\!\!\!\!\!\simeq 1\,{\rm keV}\!\left(\frac{4\,{\rm km/s}}{\sigma}\right)^{1/4}\!\!\left(\frac{50\,{\rm pc}}{r_{c}}\right)^{1/2} \, 
\end{align}
if the $\chi$ distribution is well approximated by an isothermal sphere.
For example, for the ultra-faint dwarf spheroidal galaxy Leo IV we have $m_{\chi} \gtrsim 0.7$\,keV \cite{Simon:2007dq}, although rigorously this bound is slightly weaker \cite{Boyarsky:2008ju}.  This is shown in Figure.~{\ref{fig:constraints} as the curve labeled \mbox{\emph{Phase Space}}. 

\vspace{0.04in}
\noindent \emph{Extreme Models} -- The maximum $m_{\chi}$  allowed, 
\begin{align}
m_{\chi}^{\rm max} \simeq 5 \,{\rm TeV} \!\left(\frac{10^3}{{\cal E}_r}\cdot \sqrt{\frac{\Omega_d h^2}{0.11}}\right)^{4/5}\!\!\left(N_{\rm ch}^3\cdot \frac{d_{\chi}}{4}\sqrt{\frac{\hat{g}_{r,f}}{107}} \right)^{1/5} \, ,
\label{eq:max1}
\end{align}
is  at $m_{\Lambda} \simeq 150$~TeV, at the confluence  of the decoupling, recoupling, and perturbativity constraints.  If the four-fermion condition is required instead, at $m_{\Lambda} \simeq 400$~TeV,
\begin{align}
m_{\chi}^{\rm max} \rightarrow 26.5 \, {\rm TeV}\!\left(\frac{\Omega_d h^2}{0.11}\right)^{2/5}\!\!\left(N_{\rm ch}^3\cdot \frac{d_{\chi}}{4}\sqrt{\frac{\tilde{g}_f}{107}} \right)^{1/5} \, .
\end{align}
Here, we have $\hat{g}_{r,f}\equiv \tilde{g}_{r}(\tilde{g}_{r}/\tilde{g}_{f})^{1/3}$, and for a SM thermal history $\hat{g}_{r,f} \simeq \tilde{g}_f \simeq g_f  = 106.75$.  We find viable models require \mbox{$m_{\chi} \lesssim 10\,{\rm TeV}$}.  Although $3\,{\rm keV} \lesssim m_{\chi} \lesssim 10$~TeV, viable $(m_{\chi},F_{\chi})$ regions exist over the wide range of the high-energy scale \mbox{$100~{\rm keV} \lesssim m_{\Lambda} \lesssim 10^{8}$~TeV}.

%\section{Standard Model Interactions}
\vspace{0.04in}
\noindent \emph{Standard Model Interactions} -- A key requirement of this scenario is that the HS never thermalized with the SM sector.  This limits the coupling between the HS and the SM.  The HS might interact with the SM via several channels, and each constraint is best handled individually for each particular channel and model.

We illustrate here a case where four-fermion interactions with SM fermions dominate and scale like $\sigma_{I} \sim \lambda_{I}^4/\ManS$ above an interaction scale $m_{I}$ and as $\sigma_{I} \sim G_{I}^2\ManS$ below it, where $G_{I} \equiv \lambda_{I}^2/m_{I}^2$ is an effective coupling for SM-HS interactions.  We write the interaction rate as
\begin{align}
\Gamma_{I} = \sum_{i,j \in SM} n_{i} \left(\sigmaivs+\sigmaiv \right) \equiv F_{I}^2 T^5 \, ,
\label{eq:int1}
\end{align}
when $T \lesssim m_{I}$.  Here, the sum extends over all channels that produce at least one $\chi$ particle, connect the SM to the HS, and $F_{I} \sim G_{I}$  is the thermalized coupling.  As $T_{h} < T$ we limit the HS production rate from SM particles $\Gamma_I \equiv \Gamma_{SM \rightarrow HS}$ and not the smaller rate $\Gamma_
{HS \rightarrow SM}$.  When $T \gtrsim m_{I}$ the interaction rate scales as
%\begin{align}
$\Gamma_{I} \simeq F_{I}^2 m_{I}^4 T$.
%\label{eq:int2}
%\end{align}
To determine the abundance $Y_{\chi}^{SM} \equiv n_{\chi}^{SM}/s$ of $\chi$ particles generated from SM plasma we integrate the appropriate Boltzmann equation for these processes to find
$Y_{\chi}^{\SM} = \int_{0}^{t_0} dt \,  \Gamma_{I} Y^{eq}$.
Here $Y^{eq} \equiv n^{eq}/s$ is the abundance of a species in equilibrium with the SM plasma.  The ratio $\Gamma_I/H$ peaks near $T \simeq m_{I}$ and most $\chi$ particles generated via SM-HS interactions are produced near this maximum.   For the rates above we find
\begin{align}
\frac{n_{\chi}^{\SM}}{n_{\chi}} \simeq \frac{5}{4} F_{I}^2 m_{I}^3 M_{\rm Pl}\left(\frac{m_{\chi}}{\rm keV}\right)\left(\frac{1000}{\sqrt{g_I}g_I^s}\right)\left(\frac{0.11 \cdot d_{\chi}}{\Omega_{d} h^2 \cdot 4}\right)  \, .
\end{align}
If we require $n_{\chi}^{\SM}/n_{\chi} \leq \epsilon^{\SM}$ we find the constraint
\begin{align}
\!\!\left(\frac{F_{I}m_{I}^{3/2}}{10^{-10}\, {\rm TeV}^{-1/2}}\right)\!\sqrt{\frac{m_{\chi}}{10 \, {\rm keV}}} \!\lesssim 2.6 \,\sqrt{\frac{\epsilon^{\SM}}{0.01} \frac{\sqrt{g_I}g_I^s}{1000}\frac{\Omega_{d} h^2}{0.11}\frac{4}{d_{\chi}}}
\end{align}
on $F_I m_{I}^{3/2} m_{\chi}^{1/2}$.   Note we have neglected potential resonant effects near $T \simeq m_{I}$ which may occur in some models and would lead to even more stringent constraints. 

Interactions of this type connecting the HS and SM must be \emph{extra weak}, much weaker than the Fermi interaction, with $F_{I} \ll G_{F} = 1.166 \times 10^{-5}\, {\rm GeV}^{-2}$.   If the characteristic energy scale of these interactions were $m_{I} \sim {\rm TeV}$ then typical couplings are $\lambda_{I} \lesssim 10^{-5}$ for $m_{\chi} \sim 10$~keV.  

While $\chi$ particles may be absolutely stable, extra-weak interactions satisfying this bound might allow the decay of $\chi$ particles to SM states on timescales longer than the age of the Universe.   For instance, for a three-body decay we expect a rate like $\Gamma_{\rm d} \sim G_{I}^2 m_{\chi}^5/(192\pi^3)$.  A lifetime \mbox{$\tau_{\rm d} \sim 9 \times 10^{4}\, (13.7\, {\rm Gyr})\,  (10^{-11}\, {\rm TeV^{-2}}/G_{I})^2 ({\rm MeV}/m_{\chi})^5$}, orders of magnitude longer than the age of the Universe $\sim\!\!13.7$~Gyr, is possible.  We leave the investigation of this potentially important signature for future study.

%\section{Minimal Models}
\vspace{0.04in}
\noindent \emph{Minimal Models} --
There are many possibilities, both bosons and fermions, for HHDM phenomenology. As a concrete example we have focused here, in analogy with SM neutrinos interacting via the Fermi interaction, on a model with interactions that decouple when they are well approximated by a four-fermion vertex of strength $G_{\Lambda}$.

A simple model with the right phenomenology consists of our dark-matter particle $\chi$ and another lighter particle $\psi$ ($m_{\psi} \lesssim m_{\chi}$) interacting via a spontaneously broken $U(1)$ gauge symmetry.  If coupled via a massive vector boson $U_{\mu}$ of mass $m_{U} \equiv m_{\Lambda}$ with coupling $g_{U} \equiv \lambda$, their interactions are well approximated by a four-fermion interaction of strength $G_{\Lambda} \sim g_{U}^2/m_{U}^2$ provided $m_{\chi} \ll m_{U}$ and $T_{\chi f} \ll m_{U}$.  This HS could reheat via direct couplings of $\chi$ and $\psi$ to the inflaton, or via the hidden Higgs sector associated with the broken $U(1)$.  A thermal population of $\chi$ and $\psi$ particles will be produced in the early Universe with $\chi$ accounting for most of the dark matter.   The $\psi$ particles also add to the dark matter density and, as they have a longer free-streaming length, might imprint detectable features on the matter power spectrum.

Other alternatives realizing a four-fermion model include fermions interacting via a scalar or a vector boson with more complex gauge interactions.  For instance, the interactions of $\chi$ might be a HS copy of the weak interactions of SM neutrinos.   It would be interesting to find well-motivated theories realizing the HHDM scenario.

%\section{Conclusions}\label{sec:conclusions}
\vspace{0.04in}
\noindent \emph{Summary} --
We have shown that a thermalized hidden-sector species, with interactions the decouple when relativistic, can act as cold rather than hot dark matter provided it interacts weakly enough with standard model particles to never thermalize with them in the primordial plasma.  This class of dark matter candidates can  account for all of the dark matter in the Universe and has an allowed range of masses between a few $\rm keV$ to a few $\rm TeV$.
Signatures of this mass scale may be imprinted on the small-scale power spectrum of dark-matter fluctuations or, if the dark matter decays to standard model particles on cosmological timescales, in the energy spectrum of potential decay products.

%\section*{Acknowledgements}
\vspace{0.04in}
\noindent \emph{Acknowledgements} --
This work is supported by a Natural Sciences and Engineering Research Council (NSERC) of Canada Discovery Grant.  KS would like to thank the Aspen Center for Physics, the W. M. Keck Institute for Space Studies at Caltech, and the Perimeter Institute for Theoretical Physics, where part of this work were completed, for their hospitality.

\end{document}